\documentclass[iop,rev4]{emulateapj}
\usepackage{apjfonts}
\usepackage[usenames,dvipsnames]{xcolor}
\definecolor{blue}{rgb}{0,0,1}
\usepackage[hyperfootnotes=false]{hyperref}
\usepackage{color}
\usepackage{graphicx}

\hypersetup{
  colorlinks,
  citecolor=blue,
  linkcolor=blue,
  urlcolor=blue}

\slugcomment{}

\shorttitle{Stellar Flares in CSTAR Field}
\shortauthors{Liang et al.}

\begin{document}
\title{Stellar Flares in the CSTAR Field: Results from the 2008 Data Set}
\author{En-Si Liang \altaffilmark{1}, Songhu Wang \altaffilmark{1,2,6}, Ji-Lin Zhou \altaffilmark{1}, Xu Zhou \altaffilmark{3}, Hui Zhang \altaffilmark{1}, Jiwei Xie \altaffilmark{1}, Huigen Liu \altaffilmark{1}, Lifan Wang\altaffilmark{4}, M. C. B. Ashley \altaffilmark{5}}

\altaffiltext{1}{School of Astronomy and Space Science and Key Laboratory of Modern Astronomy and Astrophysics in Ministry of Education, Nanjing University, Nanjing 210093, China; \email{zhoujl@nju.edu.cn}}
\altaffiltext{2}{Department of Astronomy, Yale University, New Haven, CT 06520, USA; songhuwang86@gmail.com}
\altaffiltext{3}{Key Laboratory of Optical Astronomy, National Astronomical Observatories, Chinese Academy of Sciences, Beijing 100012, China}
\altaffiltext{4}{Purple Mountain Observatory, Chinese Academy of Sciences, Nanjing 210008, China}
\altaffiltext{5}{School of Physics, University of New South Wales, NSW 2052, Australia}
\altaffiltext{6}{Department of Astronomy and Astrophysics, University of California at Santa Cruz, Santa Cruz, CA 95064, USA}

\begin{abstract}
The Chinese Small Telescope ARray (CSTAR) is the first Chinese astronomical instrument placed in Antarctica. It is a group of four identical, fully automatic $14.5\,\rm{cm}$ telescopes, with an field of view (FOV) of $20\,\rm{deg^2}$ centered on the South Celestial Pole. Placed at Antarctic Dome A, CSTAR is designed to provide high-cadence photometry for site monitoring and variable sources detection. During the 2008 observing season, CSTAR has taken high-precision photometric data for 18,145 stars around the South Celestial Pole. At $i\,=\,7.5$ and $12$, the photometric precision reaches $\sim 8$ mmag and $\sim 30$ mmag with a cadence of 20s or 30s, respectively. Using robust detection method, we have found 15 stellar flares on 13 sources, including two classified variables. We have also found a linear relation between the decay times and the total durations of the stellar flares. The details of all detected flares along with their stellar properties are presented in this work.
\end{abstract}

\keywords{methods: data analysis --- stars: activity --- stars: flare --- surveys --- techniques: photometric --- techniques: statistics}

\section{Introduction}

 Stellar flares are very common astronomical phenomena. Basically, they are large explosions that happen on the stellar surface, with durations from a few minutes to several hours, and energy levels up to $10^{38}$ erg for some F or G type stars \citep{Schaefer2000}. \citet{Parker1963} explains the physical mechanism of stellar flares as reconnection and annihilation of the magnetic fields in the coronal region, which releases massive magnetic energy. Many modifications have been applied to this theory in the following years (e.g. Sweet 1969, Heyvaerts et al. 1977, Shibata \& Magara 2011), and theories on potential flaring triggers such as a close-in companion have also been proposed (e.g. Rubenstein \& Schaefer 2000). Studying the mechanism of stellar flares provides us a way to better understand the magnetic activities on the stellar surface, while probing the relations between the statistical features of stellar flares and the properties of their host stars enables us to better understand the magnetic evolution of different types of stars. Moreover, as the search for habitable exoplanets gradually heats, the impact of massive flare eruptions on the stellar habitable zone is gaining more attention.

Since stellar flare events occur with no known signs, simultaneous and continuous observations of large number of stars help increase the chance of detecting stellar flares. Therefore, wide-field, continuous and precise photometric surveys offer higher flare-detection chance. The recent launch of the $Kepler$ Spacecraft made great progress on observing stellar luminosity change with extremely high precision \citep{Borucki2010}. Its data helped bring out some fascinating yet puzzling flare cases, such as stars originally thought to be able to generate only small flares turn out to have flares much larger (e.g. Maehara et al.  2012; Balona 2012), which shows that our understanding of the mechanisms of stellar flares is not complete. Back to the ground, besides global cooperated longitude-distributed observing programs such as HATNet \citep{Bakos2004} and HATSouth \citep{Bakos2013}, Antarctic Dome A provides an alternative for continuous stellar photometric observation. Observing qualities on Dome A are thought to be one of the best on Earth (Burton et al. 2010; Saunders et al. 2009; Zhou et al. 2010a).

To utilize this extraordinary astronomical observing conditions of Dome A, in January 2008, the CSTAR telescope was shipped and installed at Dome A. During the same year, it acquired almost 0.3 million scientific qualified frames in the $i$-band through 4 months of nearly non-stop observation. \citet{Zhou2010a} released the first version of the photometric catalog in 2010, and it has been corrected for additional systematic errors by \citet{Wang2012}, \citet{Meng2013} and \citet{Wang2014a}.

Various works have been done with the CSTAR data thus far (e.g. variable sources: Wang et al. 2011, Huang et al. 2015, Wang et al. 2015, Zong et al. 2015; transiting exoplanets: Wang et al. 2014b; binaries: Yang et al. 2015). Yet there are still non-periodic transient events undiscovered, such as stellar flares. Therefore, in this work, we develop a custom and robust method to search for stellar flares in the CSTAR data set, and study the properties of the detected flares.

%The structure of the paper is as follows. A brief introduction of the CSTAR instrument, observation and previous data reduction is presented in Section 2. In Section 3, we describe the methods we applied to detect stellar flares from the CSTAR data. In Section 4, we present the detected flares along with their detailed properties. Lastly, we briefly summarize our work and discuss future work prospects in Section 5.

\section{Instrument, Observations and Previous Data Processing}

\subsection{Instrument}

The full instrumental details of the CSTAR system are described in \citet{Yuan2008} and \citet{Zhou2010b}. Here we briefly describe the key parameters of the installation related to our work. Controlled from the PLATO unmanned observatory \citep{Lawrence2009,Yang2009}, the CSTAR facility consists of four co-aligned Schmidt-Cassegrain telescopes with the same $4.5^\circ \times 4.5^\circ$ FOV pointed at the South Celestial Pole. Three of the four telescopes are equipped with a different filter in the SDSS bands: \textit{r, g, i}, and the other one without any. For each telescope, a $1K \times 1K$ Andor DV 435 frame transfer CCD array is coupled with an entrance pupil with a diameter of 145 $\rm{mm}$, which has an effective aperture of 100 $\rm{mm}$. The corresponding plate-scale is 15 arcsec pixel$^{-1}$.

\subsection{Observations}
CSTAR was successfully installed at Dome A in January, 2008. It went into fully operational mode for the next four winters, from 2008 to 2012. The light curves we analyzed in this work were all from the 2008 observing season, March 4 to August 8, during which about 0.3 million frames with exposure time of 20s or 30s were taken for 18,145 stars. For the details of the CSTAR observations, we refer the readers to \citet{Zhou2010a}.

\subsection{Previous Data Reductions}
In order to achieve mmag precision for bright CSTAR objects, a custom reduction and calibration pipeline were developed, as described in detail in \citet{Zhou2010a} and \citet{Wang2012,Wang2014a}. We briefly describe the data reduction procedure here.

After the correction of bias and flat field, aperture photometry was applied on all frames taken by the CSTAR telescope. The instrumental magnitudes of CSTAR was then converted to the SDSS $i$ using 48 local bright reference stars. The first version of the CSTAR catalog was then constructed, as described in \citet{Zhou2010a}.

To detect stellar flares, further improvements in photometric precision are essential. The correction of remaining systematic error that is closely related to detecting stellar flares is briefly described here.

Non-uniform extinction caused by the atmosphere across the large FOV of CSTAR, especially in poor weather conditions, cannot be ignored, when photometric precision reaches mmag. This systematic error can be modeled and corrected by comparing each frame to a master frame, which is generated using the frames under the best observing conditions. Detailed description on how this correction was made is available in \citet{Wang2012}.

After the correction described above, the typical photometric precision of the light curves we used in this work reaches 8 mmag. A precision of 4 mmag can be achieved if the diurnal effect (periodical contamination) is corrected. See \citet{Wang2014a} for more details.

\section{Flare Detection}
%In this section we focus on introducing the custom methods we developed for flare detection in the CSTAR data set. We start by briefly reviewing the photometric precision of the light curves, then we go into describing the robust methods we developed to search for stellar flares, finally we discuss the possible contaminations that may generate flare-like signals and the corresponding validation procedure. 

\subsection{Photometric Precision}
\begin{figure}
\plotone{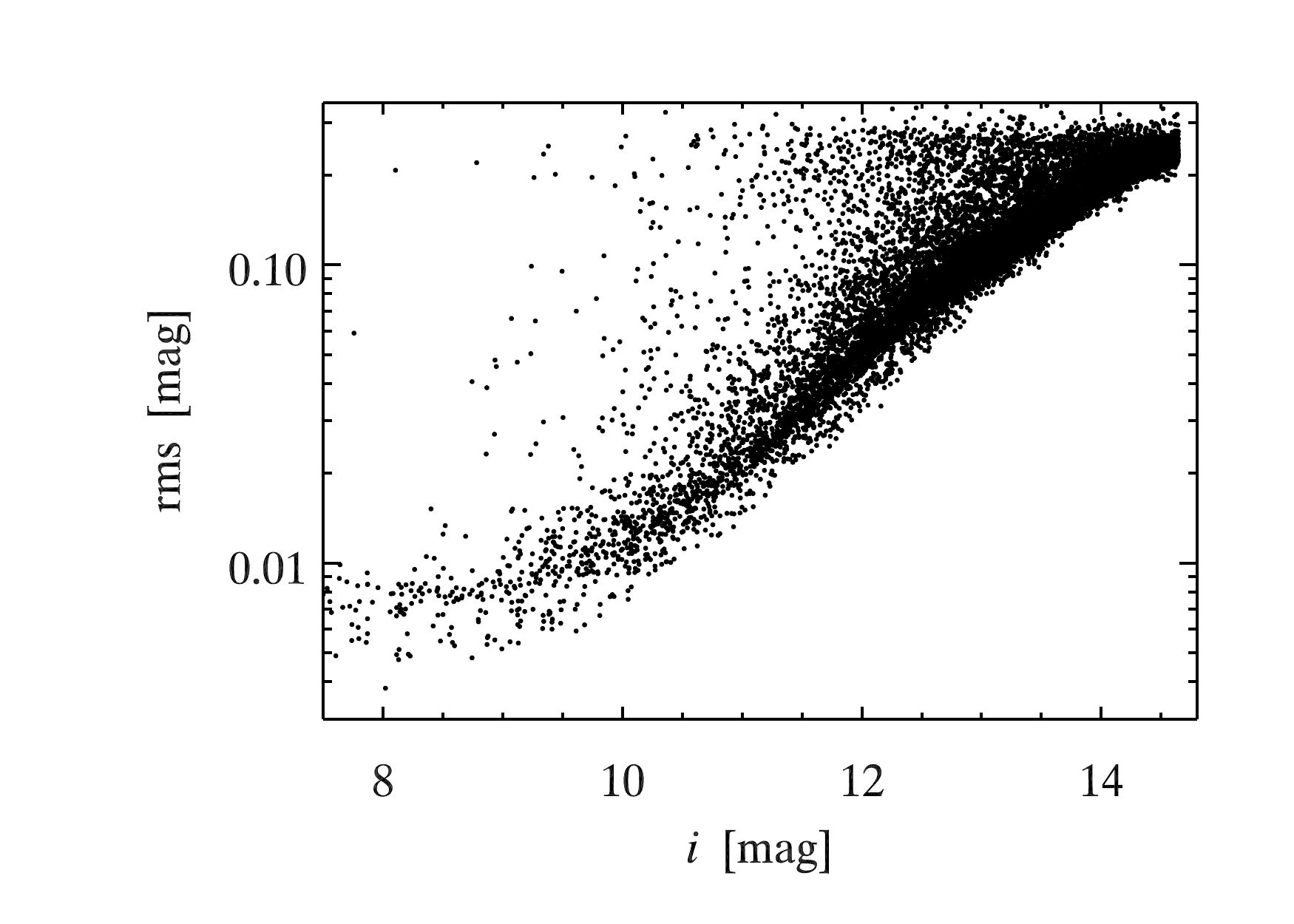}
\caption{CSTAR photometric quality of the 2008 data set. To best illustrate the photometric precision of the light curves, the y-axis is drawn in log scale. At $i=7.5$, the CSTAR data precision reaches $\sim$ 8 mmag. 
\label{fig1}}
\end{figure}

In our work, saturated stars are first filtered out since their photometric measurement are not precise. For stars with SDSS $i$ magnitudes just below the saturation limit ($i$=7.5), typical photometric precision reaches $\sim$ 8 mmag at 20s or 30s cadence. As the magnitude of the stars rises, typical photometric precision worsens to $\sim$ 30 mmag at $i$=12. We show in Figure \ref{fig1} the rms distribution plotted against the $i$-band magnitudes of the light curves, where the y axis is in log scale.

\subsection{Flare Detection Method}
\begin{figure}
\epsscale{0.85}
\plotone{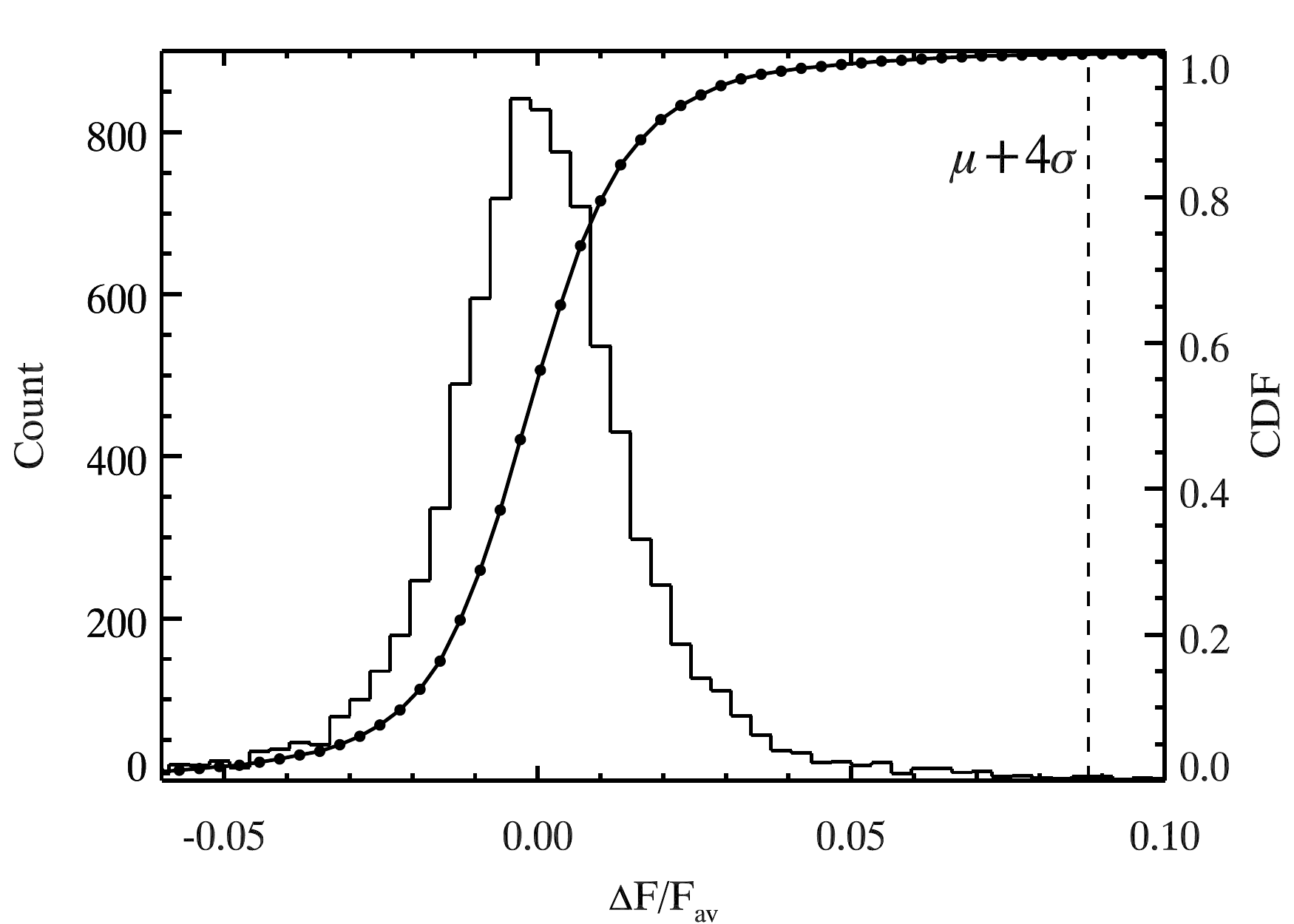}
\caption{Distribution of flux difference between consecutive bins of flaring variable 2MASS J174728.657-884609.43. The histogram exhibits a Gaussian distribution centered around zero. We define any flux difference greater than $\mu+4\sigma$ (threshold) of the distribution as a flare candidate. 
\label{fig2}}
\end{figure}

For a star in its quiescent state, the observed flux variations shows up on light curves in a stable and random pattern. The flux difference between consecutive points on light curves follows a Gaussian distribution. When a flare event occurs, a corresponding flux difference anomaly appears on the light curve. We define any flux difference greater than $\mu+4\sigma$ (Figure \ref{fig2}) of the distribution as a flare candidate. However, since the flux-rising period of a flare covers a range from several minutes to almost an hour, which is larger than the CSTAR working cadence, we highlight the gap between quiescence and peak by binning the data points with different intervals (3, 5, 10, 20 and 30, in unit of minute).

\begin{figure}
\epsscale{0.85}
\plotone{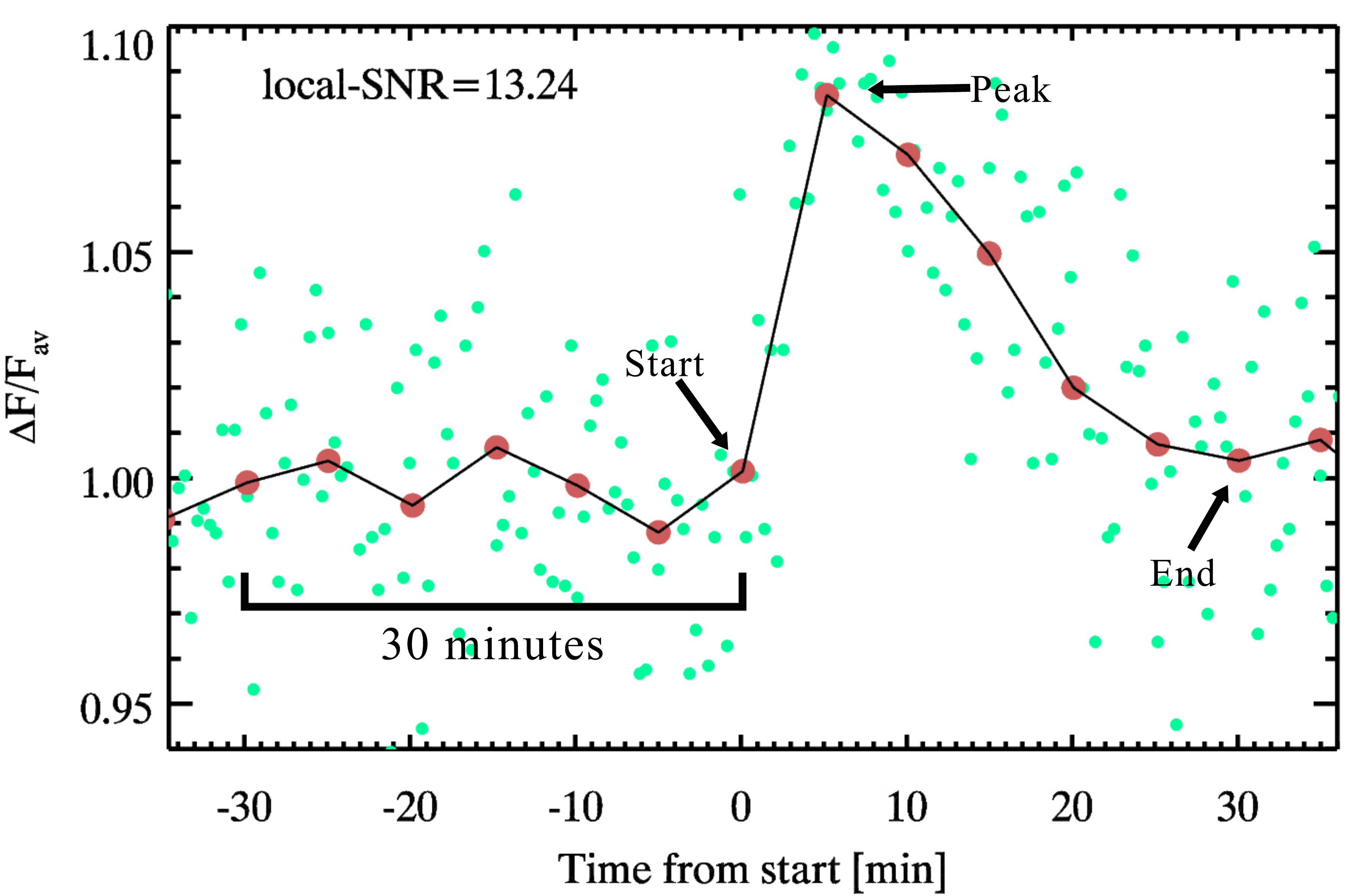}
\caption{ Demonstration of the local-SNR parameter and the start, peak, and end points using detected flaring source 2MASS J031821.47-881506.29. We calculate the ratio between the flare amplitude and the standard deviation of the flux before the flare candidate. The segments we use to compute the standard deviation in this case is of 30 minutes long, which is 6 times the bin interval we applied. If the local-SNR is greater than 3, this flare event will be further validated. In this case, the local-SNR value is 13.24, much larger than 3.
\label{fig3}}
\end{figure}

Remaining systematic errors, heavy extinction caused by occasional bad weathers could produce lesser quality observation segments. These segments, usually with large photometric dispersion, could result in some of the low-quality data points being selected as flare candidates. The local-signal-to-noise-ratio (local-SNR) parameter is hence introduced to ensure that the selected flux difference anomaly is prominent comparing to the stellar quiescent state. As shown in Figure \ref{fig3}, the local-SNR parameter calculates the ratio of the amplitude of a flare candidate to the standard deviation of the flux variation before the flare candidate. Any anomalies with an local-SNR value greater than 3 will be further validated.

\subsection{Validation Process}
Although the local-SNR parameter excluded a large proportion of the false positive flare events, our result inevitably suffers from remaining systematic errors. Thus, further validation procedures are required to eliminate remaining false positive signals.

\begin{itemize}
\item Ghost image contamination. Ghost image is caused by the reflection of light between different mirrors installed in the CSTAR telescope. Since CSTAR is a fixed mount telescope pointed to the South Celestial Pole, diurnal motion causes the stellar images to rotate clockwise on the CCD plane. Thus, the reflected image, which is called ghost image, rotates counter clockwise on the CCD plane. When a ghost image happens to overlap another star, the luminosity of the contaminated star might increase significantly \citep{Meng2013}. This kind of luminosity increase could disguise as a flare signal with almost 1-day periodicity, but the amplitude of each peak is not always the same, mainly due to the small difference in overlapping area caused by the constant shifting of the ice layer on which CSTAR was mounted. Thus, we dispose of any flare candidates with significant peaking occurring at around ($T_{\rm peak}\pm1$ sideral day) or ($T_{\rm peak}\pm2$ sideral days), where $T_{\rm peak}$ is the flare candidate peaking time. Moreover, on the light curves, false positives caused by ghost image contamination demonstrate near axial symmetry about their peaks, which enables us to remove any remaining false positives with near axial symmetry by comparing their up and down durations.

\item Neighboring star contamination. Flare-like signals may be detected on originally quiescent sources if contamination by nearby flaring stars occurs. Although the total number of false positives caused by neighboring star contamination is rather small, we eliminate these ambiguous signals if the light curves of their neighboring sources are found to display simultaneous rise, for conservative reasons.

\end{itemize}

In addition, to avoid any uncertainty, each remaining flare candidate is further inspected by eye. We manually make sure that most of the remaining possible contamination sources, such as hot pixel, cosmic ray, random satellite track are all taken into account. After thorough visual inspection, 15 reliable flare events on 13 stars are discovered in the CSTAR 2008 observing data.

\section{Result and Discussion}

\subsection{Result of Flare Searching Process}
We use 18,145 light curves from the CSTAR 2008 observing data set to search for stellar flares, with the brightest non-saturated object up to $i=7.5$ and faintest object down to $i=14.8$. Using robust and data-customized method, we find 15 flare events on 13 CSTAR sources (Figure \ref{fig4a}). Notable characteristics of all 15 flare events along with their stellar properties are listed in Table \ref{tab2}. The identifier is based on the stellar coordinates from the 2MASS system \citep{Skrutskie2006}. In addition, all the light curves of the detected flaring sources are available online\setcounter{footnote}{4}\footnote{http://explore.china-vo.org/}.

\begin{figure*}
\epsscale{0.85}
\plotone{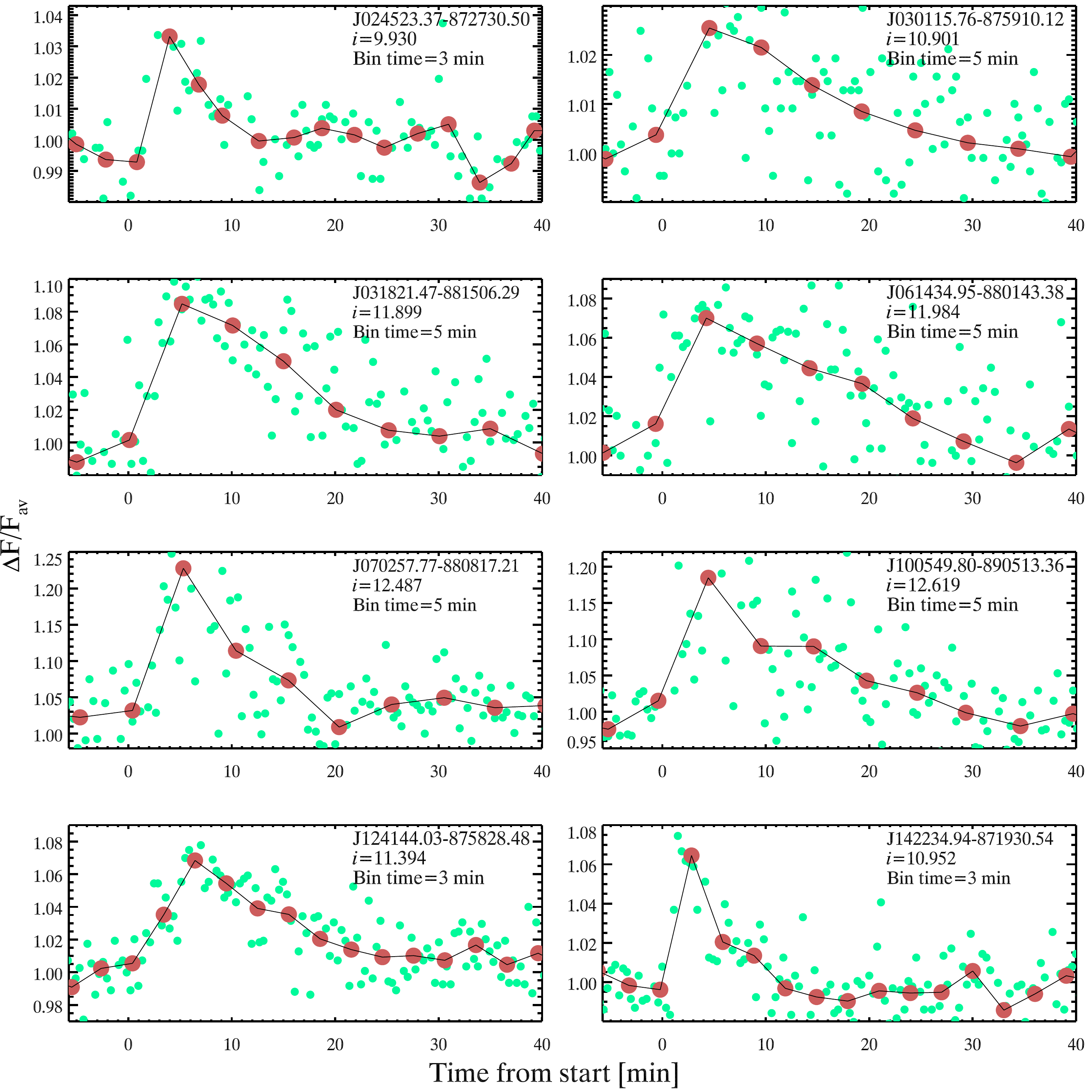}
\caption{$i$-band photometric data for 15 flare events discovered in the CSTAR 2008 data set. The 2MASS identifiers and mean magnitudes of the sources on which the flare events occurred are shown on the upper right corner. Green data points are original unbinned data, connected red data points represent binned data using different bin times as shown on the upper right corner. The results are presented in the same time scale and arranged in order of increasing right ascension coordinates except for the one with a duration of 260 minutes, which is placed at the end.} 
\label{fig4a}
\end{figure*}

\renewcommand{\thefigure}{\arabic{figure}}
\addtocounter{figure}{-1}

\begin{figure*}
\epsscale{0.85}
\plotone{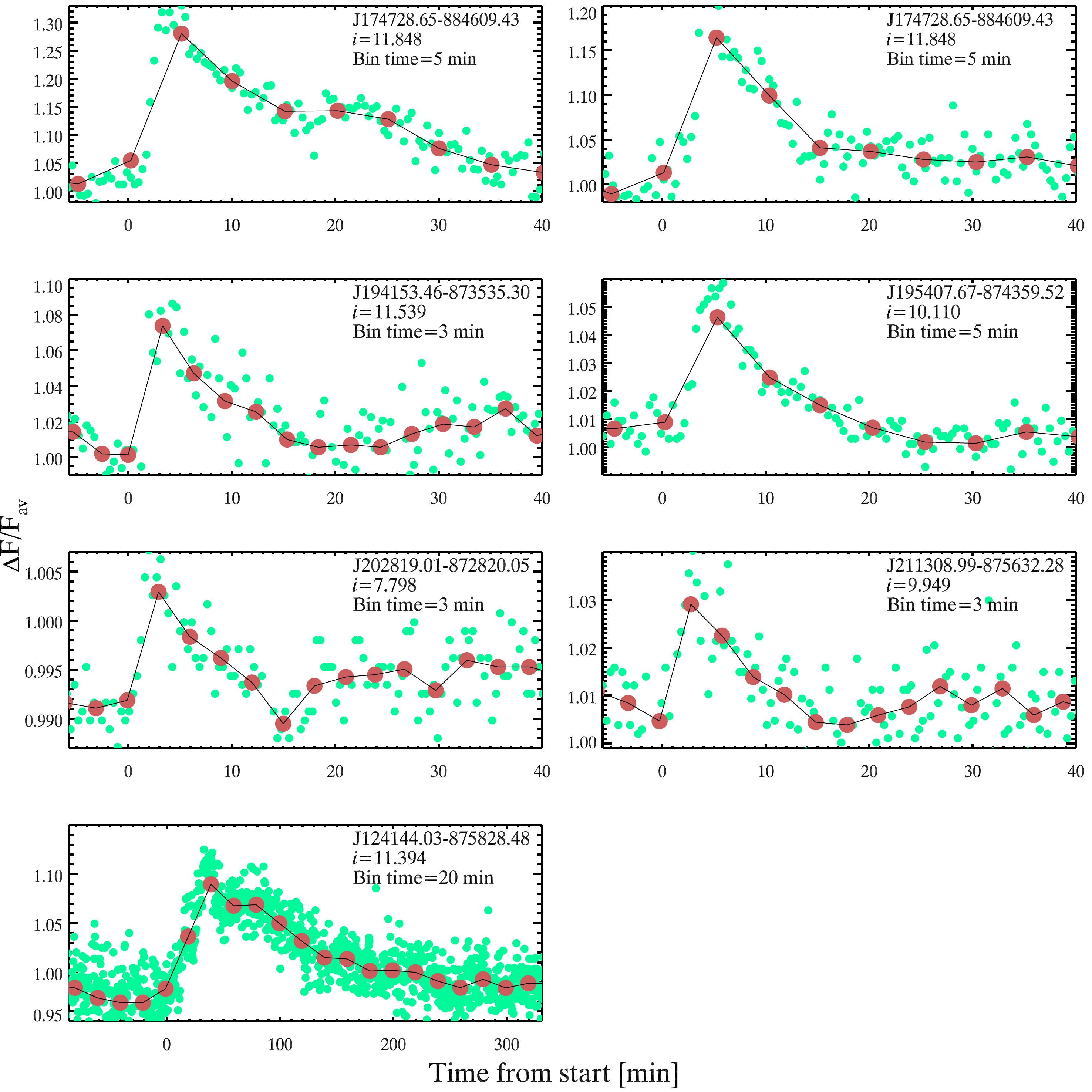}
\caption{Continued
\label{fig4b}}
\end{figure*}

\subsection{Properties of Discovered Stellar Flares}
\begin{figure}
\epsscale{0.85}
\plotone{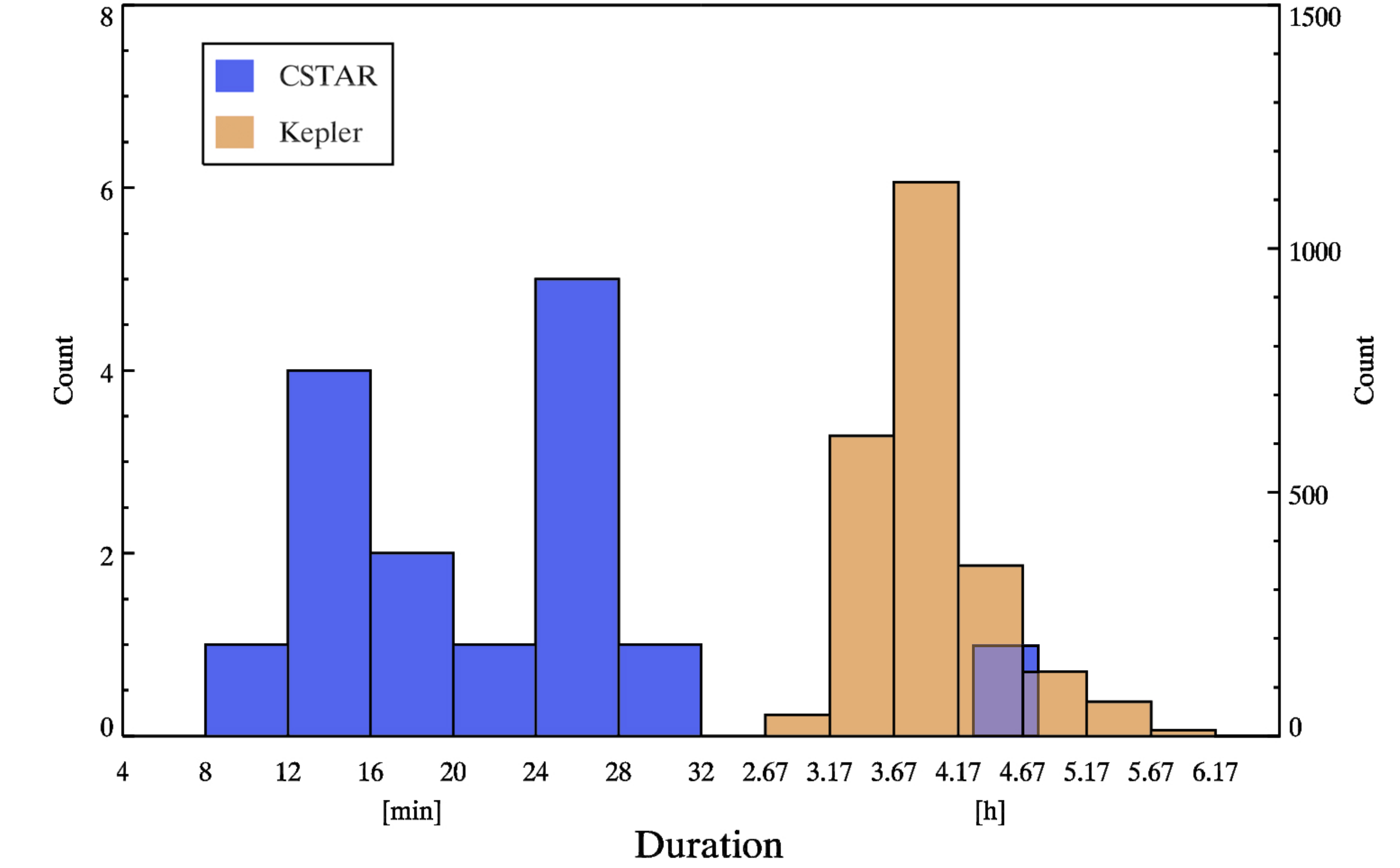}
\caption{Distribution of flare durations calculated using both $Kepler$ data and CSTAR data. Orange bars draw the duration distribution of $Kepler$ flares \citep{Walkowicz2011}, and blue bars represent CSTAR result. The CSTAR findings concentrate in the short duration region, mostly shorter than one hour, with one 260-minute exception pointed out in the right panel.
\label{fig5}}
\end{figure}

One of the advantages of using the CSTAR observing data to search for stellar flares is that the high working cadence of the CSTAR telescope provides us with an opportunity to detect short duration stellar flares. As shown in Figure \ref{fig5}, the durations of the stellar flares found in the CSTAR data set mainly concentrate in the zone between 10 minutes and 40 minutes, with an exception that goes to 260 minutes. However, hampered by its working cadence, flares found in the $Kepler$ long cadence data set \citep{Walkowicz2011} are always longer than one hour.

We define the amplitude ($A$) of a stellar flare as follows,
\begin{equation}
A=\frac{\Delta F_{\rm max}}{F_{\rm av}}\times 100\%,
\end{equation} 
where $\Delta F_{\rm max}$ is the flux difference between the highest point and the neighboring quiescent segment ($F_{\rm av}$) of the flare. The amplitudes we calculated cover a very wide range, from as small as $1\%$ to as large as $27\%$. Small amplitude stellar flares are found thanks to the precise photometric ability of CSTAR that reaches 8 mmag at brightest level $i$ magnitude.

As shown in Figure 4, almost all 15 flare events undergo a fast rising period and followed by an slower exponential decaying pattern, which is in accordance with previous studies (e.g. Moffett \& Bopp 1976; Kowalski et al. 2013). To define the ratio of time between energy injection (rise period) and release (decay period), we introduce a new parameter, skewness ($\kappa$), which is in the form of
\begin{equation}
\kappa=\frac{T_{\rm des}}{T_{\rm asc}},
\end{equation} 
where $T_{\rm asc}$, measured from the prior point of the detected flux anomaly to the peak of the flare, denotes the time of the rise period; similarly, $T_{\rm des}$, represents the time of the exponential decay period, which is the time from the peak to the first point after the peak that is below $F_{\rm b}+\sigma _{\rm b}$, where $F_{\rm b}$ is the average flux and $\sigma _{\rm b}$ is the standard deviation of the same segment that we used to calculate the local-SNR parameter. We note that the points we mentioned above are from binned light curves, which have a higher signal-to-noise ratio. The fifth column of Table \ref{tab2} lists the skewness of all the 15 flare events that we detected.

\begin{figure}
\epsscale{0.85}
\plotone{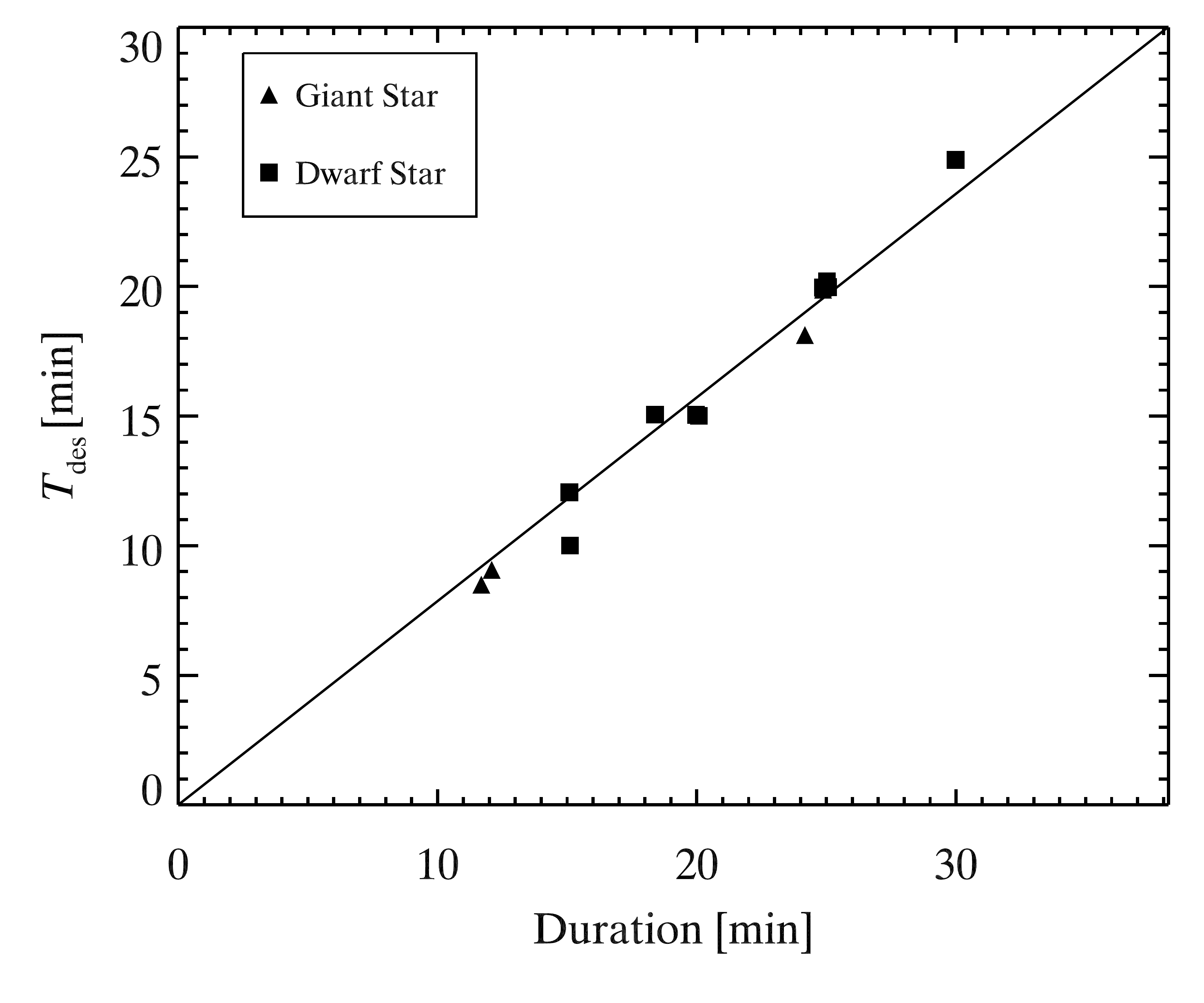}
\caption{The decay time of each flare plotted against their corresponding flaring duration. An obvious linear relation shows in this figure. Note that the 260-minute flare event lies closely to this linear line as well, but was omitted while plotting this graph to emphasize the short duration section.
\label{fig6}}
\end{figure}

Although the dispersion of skewness seems quite large, in Figure \ref{fig6}, we find a clear linear relation between the decay time ($T_{\rm des}$) and its corresponding total flare duration. This relation is surprisingly obvious even using such a small sized sample. The fitted formula is in the form of
\begin{equation}
T_{\rm des}=0.78(6)\times (T_{\rm asc}+T_{\rm des}).
\end{equation}
Using simple mathematical transformation, the ratio between the two stages can be written as
\begin{equation}
\frac{T_{\rm des}}{T_{\rm asc}}=3.67.
\end{equation}
The number in parentheses denotes the error in the final digit of the fitted coefficient. 

On the whole, as can be seen from the equations, the skewness values of short duration flares that we detected have a typical value of 3.67, which means the dissipation process usually takes 3.7 more times than the energy burst process. Larger sample size, however, is needed to further investigate this value and to fully comprehend the underlying physical implications. Once confirmed, this relation may help reveal the underlying correlation between the burst and dissipation of the magnetic energy stored on the surface of the stars.

To maximize the information about the 13 flaring sources, we utilize the stellar data with the VizieR database \citep{Ochsenbein2000}. The spectral types and luminosity classes of these sources are of our main concern, since the intrinsic properties of different classes of stars could be very unlike, resulting in dissimilar flaring mechanisms. From the VizieR database, we extract these two properties, if available, and list in Table \ref{tab2} with asterisks.

For those flaring sources whose properties are not yet known, we calculate their $T_{\rm eff}$, spectral types and luminosity classes with their $JHKBV$ magnitudes and proper motions. We note that since the CSTAR sources are not accurately registered astronomically \citep{Wang2015}, we match the detected CSTAR sources with their 2MASS counterparts by looking for sources with similar $i$-band magnitudes in a $2' \times 2'$ boxsize which only yields one matching result. The $JHK$ magnitudes of these flaring sources are then extracted from the 2MASS catalog \citep{Skrutskie2006}.

\begin{figure}
\epsscale{0.85}
\plotone{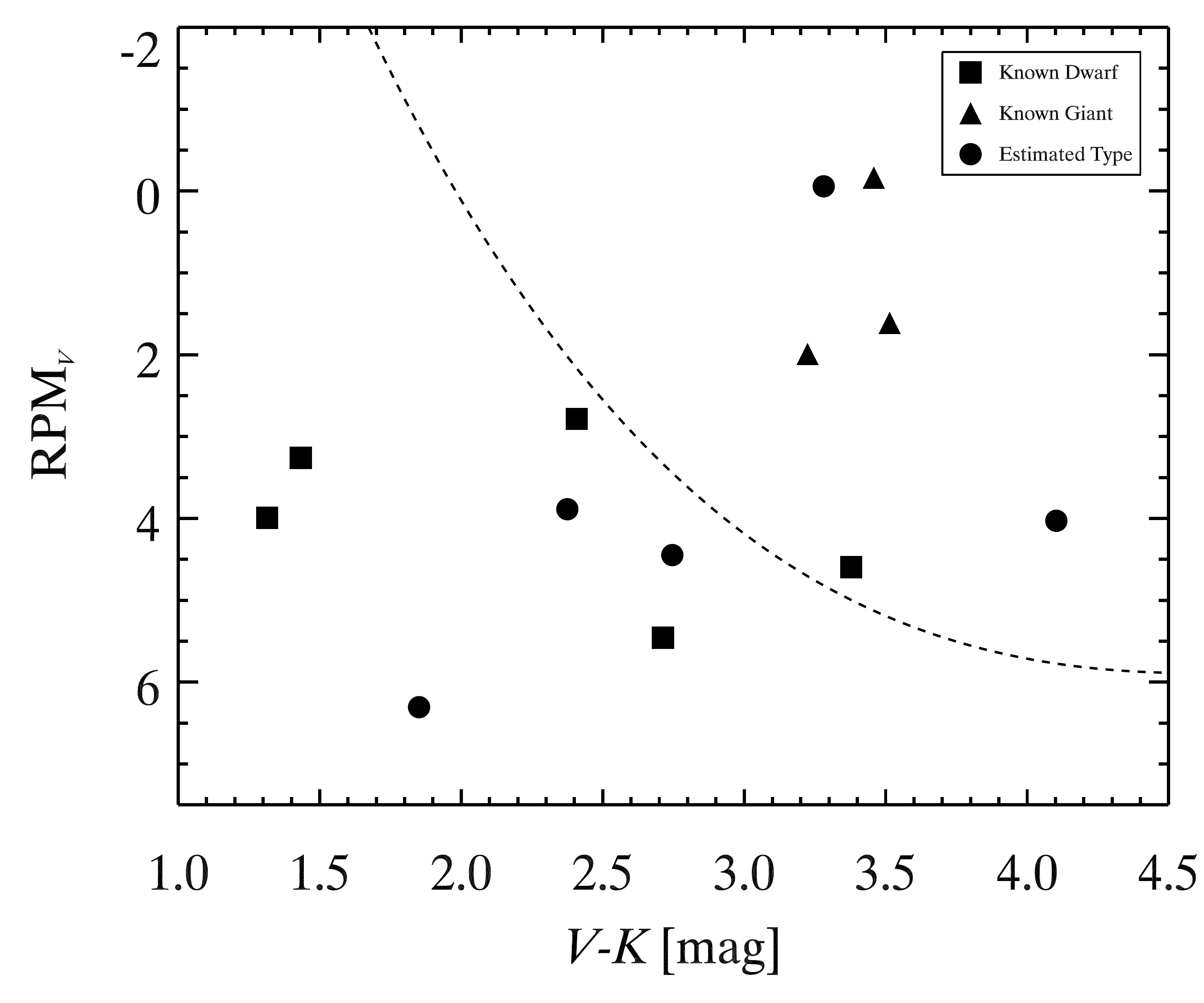}
\caption{ $V - K$ against reduced proper motion (RPM$_{V}$) diagram for the flaring sources in the CSTAR field. A polynomial dashed line (Clarkson et al. 2007) shows the boundary of the giant/dwarf separation. Dwarfs tend to concentrate on the lower left corner of the plot, while giants are more likely found on the upper right corner. Squares represent previously known dwarf stars, triangles represent known giant stars and circles represent those stars with unknown luminosity class. Most stars with known classes are well separated by the polynomial dashed line, except for one source lying on the boundary. 
\label{fig7}}
\end{figure}

Making use of the accurately matched 2MASS coordinates, we obtain the $BV$ magnitudes and proper motions of the flaring sources from the UCAC4 catalog \citep{Zacha2013}, and the PPMXL catalog \citep{Roeser2010}. Using the information obtained, we roughly separate main-sequence dwarfs from giants \citep{Clarkson2007,Hartman2011,Street2007} based on their $V - K$ color indices and RPM-reduced proper motion \citep{Luyten1922}. The ${\rm RPM}_V$ value we adopt here is calculated as
\begin{equation}
{\rm RPM}_V=V+5{\rm log}_{10}(\mu/1000),
\end{equation}
where $\mu$ is the stellar proper motion in units of mas yr$^{-1}$. Figure \ref{fig7} shows the resulting ${\rm RPM}_V$ vs ($V - K$) plot, the polynomial dashed line taken from \citet{Clarkson2007} draws a boundary between giants and main-sequence dwarfs. In this figure we plot all 13 flaring sources including 8 stars with known luminosity classes. Of all 8 known stars, 7 of them are well separated by the polynomial line, with one exception lying on the boundary, indicating that the luminosity classes of the stars can be well estimated. To estimate the unknown spectral types, we use the relation suggested by \citet{Bessell1988}. The required color indices are calculated from the catalogued $JHKBV$ magnitudes. We adopt the relations provided by \citet{Flower1996} and \citet{Torres2010} to calculate the unknown effective temperatures of the stars, required luminosity classes are from our calculation mentioned above and the VizieR database. Properties yielded from our calculations are listed in Table \ref{tab2} without asterisks.

Previous theories consider low mass late spectral type to be the most common types of stars that can generate large amplitude stellar flares due to their very deep convection zone \citep{Garcia2000}. Yet in our findings, we find a flare event on a main-sequence F4 type star (2MASS J100549.80-890513.36) with an amplitude reaching almost 20\%. Although enhanced magnetic activity have been confirmed on some late-F to M main-sequence stars \citep{Audard2000}, large amplitude flare events are mostly recorded during observations focusing on late-type stars. To further understand the various origins of stellar flares, more non-late-type flaring cases are required. Thus, further flare detection studies should tilt to F type or G type stars since their flaring mechanism may deviate from those of the late type flaring stars.

Lastly, we perform an cross-correlation with the existing variable catalogs \citep{Wang2015} and the $Rosat$ X-Ray Source Catalog \citep{Voges1999,Voges2000} to determine whether any flaring sources are variables or active X-ray emitters. This examination yield two variables and one X-ray active source. The two variables are of the type rotating ellipsoidal (ELL) and BY Draconis (BY Dra), the latter one being an active X-ray emitter as well. The last two columns of Table \ref{tab2} show the $Rosat$ status and variable status.

\subsection{Flare Events on 2MASS J174728.65-884609.43}
We discuss below two specific cases of flare events on a single star to provide the reader a taste of the data.

2MASS J174728.65-884609.43 was originally identified as a M Dwarf by \citet{Riaz2006}. \citet{Wang2011} first discovered its $i$-band variability and \citet{Wang2015} classified this star as a BY Draconis type rotating variable. Our findings reclassify 2MASS J174728.65-884609.43 as a UV Ceti star \citep{Luyten1922}. We find two flare events with very large amplitudes between a 24-day interval on this source, with the first one being a complex flare event during which two secondary events occurred, as can be seen in Figure \ref{fig4b}.

\begin{figure}[!hb]
\epsscale{0.85}
\plotone{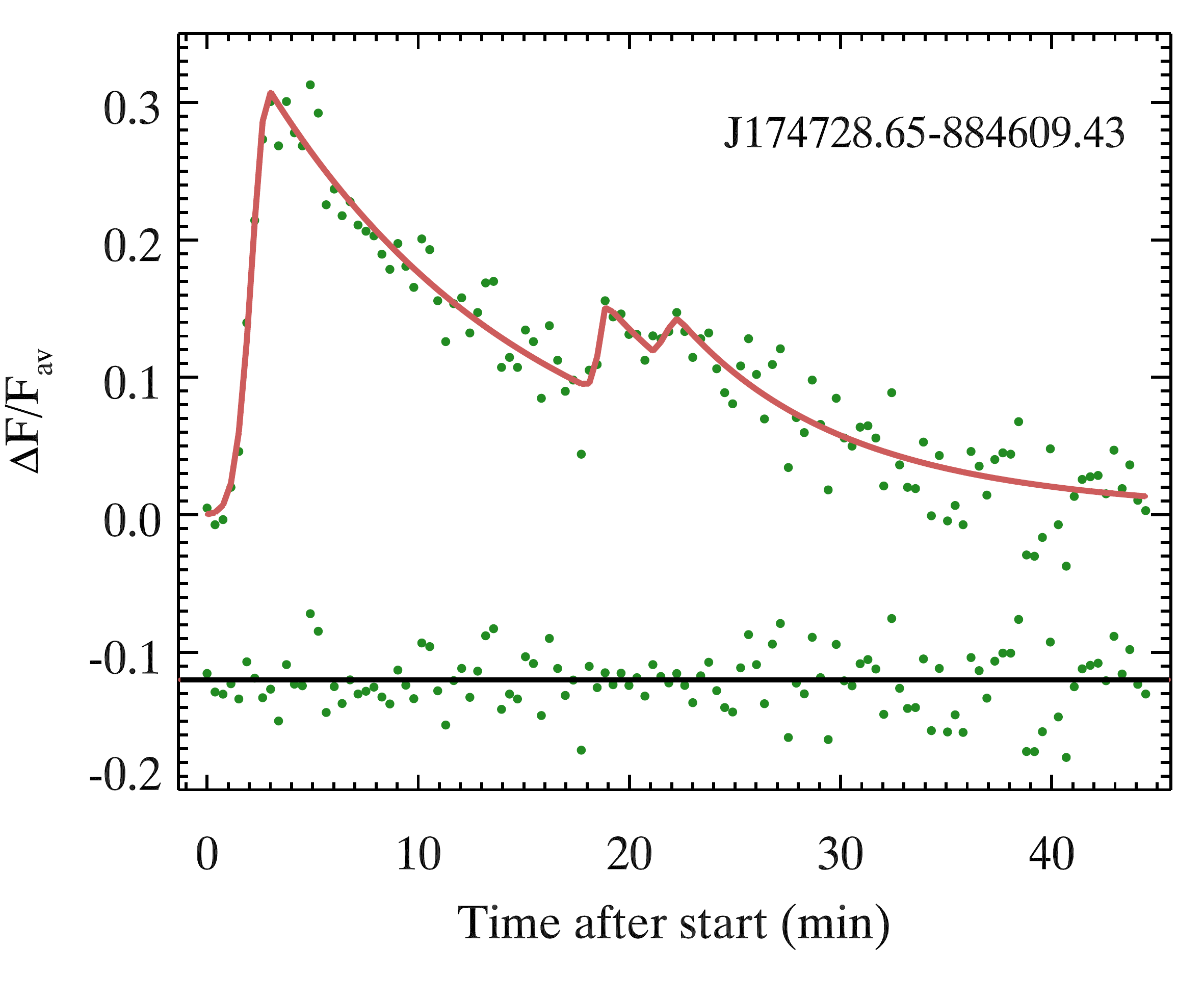}
\caption{Fitted flare model drawn upon original data. Red curve denotes the light curve we generated using models from \citet{Pitkin2014}, green points are from the original light curve. Difference between the original data and simulated data is shown on the lower part of the figure.
\label{fig8}}
\end{figure}

To reveal the true properties of the primary flare in this case, we use the flare model in \citet{Pitkin2014} to find out the duration. The model is in the form of 
\begin{equation}
F_{\rm flare}=A_{\rm 0}\{ 
\begin{array}{l}
e^{- \lambda _{\rm r} (t-T_{\rm peak})^{2}} \ \ \ \ \ \     {\rm if}\  {t<T_{\rm peak}}\\
e^{- \lambda _{\rm d} (t-T_{\rm peak})}  \ \ \ \ \ \ \    {\rm if}\  {t>T_{\rm peak}}\\
\end{array} 
\end{equation}
where $F_{\rm flare}$ is the stellar flux, $A_{\rm 0}$ is the fitted amplitude of the stellar flare, $T_{\rm peak}$ is the peak time, $\lambda _{\rm r}$ is the rising constant and $\lambda _{\rm d}$ is the exponential decay constant. These parameters are all fitted during our fitting process, and the best-fit model happens when using two successive secondary events. A detailed table of the results of the fitting process is shown in Table \ref{tab1}, and the fitted complex flare is presented in Figure \ref{fig8}.

\setlength{\tabcolsep}{2.0pt}
\begin{deluxetable}{cccccc}
\tablecaption{Fitting Results of Complex Flare Event detected on 2MASS J174728.65-884609.43}
\tablewidth{0pt}
\tablehead{
\colhead{}     &  \colhead{$T_{\rm start}$}     &  \colhead{$T_{\rm peak}$} & \colhead{$A$} & \colhead{$\lambda_{\rm r}$} & \colhead{$\lambda_{\rm d}$}  \\
\colhead{}     &  \colhead{(min)}               &  \colhead{(min)}     &  &  & \\    
}
\startdata
Primary flare & 0.00 & 2.94 & 31.1\% & 0.791 & 0.080 \\
Secondary flare 1 & 14.79 & 16.00 & 6.6\% & 4.343 & 0.162 \\
Secondary flare 2 & 18.19 & 19.41 & 3.8\% & 1.611 & 0.164 \\
\enddata
\label{tab1}
\tablenotetext{}{Note. This table shows the fitted parameters of the complex flare event detected on 2MASS J174728.65-884609.43, where $T_{\rm start}$ and $T_{\rm peak}$ are the fitted start time and peak time of each flare event; $A$, $\lambda _{\rm r}$ and $\lambda _{\rm d}$ are the fitted amplitude, rising parameter and decay parameter of each flare event.}
\end{deluxetable}

The high-precision photometry and high-cadence working mode of CSTAR provides us great details in the whole process of this event. After subtracting the two secondary events from the primary flare, we see it takes only minutes for the primary flare to reach its maximum output, and nearly 5 times longer to return to its quiescent stage. The total time elapsed is almost 30 minutes. The second flare event occurs on BJD 2454597.241, which has a duration of only 15 minutes and a skewness of 2.0. The decaying speed of this events is apparently much greater than the previous found flare. The cross-correlation with the $Rosat$ X-Ray Source Catalog also identifies this star as an X-Ray active source. Since 2MASS J174728.65-884609.43 is an active BY Dra class dwarf, its flare events release energy from X-ray to radiowave \citep{Pettersen1989}. Simultaneous multi-band observation data can help better depict flare properties, if available.

\section{Conclusion}
The CSTAR telescope, being the first Chinese astronomical instrument installed at Dome A, Antarctica, provided us with a valuable data set for detailed study of transient astronomical phenomena. About one hundred days of continuous, high-cadence photometric observation enables us to find many short duration flares, and to study the properties of flares on light curves. We create a pipeline suitable for detecting stellar flares in the CSTAR data set, with which we find 15 flare events on 13 sources, among 18,145 stars. We also provide detailed properties of the detected flares and corresponding stellar properties.

We hope our result will be a good supplementary set for the $Kepler$ flare sample, since our findings mainly focus on short duration flares. 14 out of all 15 detected flares concentrate in the 10 minutes to 40 minutes zone, with one exception overlapping with the $Kepler$ duration range. Flare amplitude, on the other hand, ranges from 1\% to 27\%, covering a rather wide interval.

Here in our work we define a new parameter, skewness, to depict the outline of flare events on light curves by calculating the ratio between up and down time. The calculated skewness, which falls in the interval between 2.00 and 5.50, has a rather large dispersion. However, a clear linear relation between flare decay time and total duration is found, where the fitted coefficients correspond to a skewness of 3.67. Once validated by larger sample of flares, the physical comprehension of this particular value needs further study with both theoretical stellar physics and MHD simulations.

We have also detected flares on a BY Dra type variable 2MASS J174728.65-884609.43 and an ELL type variable 2MASS J124144.03-875828.48, these two discovered variable now fall into a new category called UV Ceti. Meanwhile, 2MASS J174728.65-884609.43 has been discovered by the $Rosat$ telescope to be an active X-ray emitter. Subsequent X-ray flux increase can be studied in detail.

The detection of stellar flares in the CSTAR field, along with previous scientific results such as exoplanets, binaries and variables has already shown the power of CSTAR. Its complete scientific value will be truly uncovered when the $i$-band data of 2008 observing season is combined with multi-band photometric data collected in the following years.

As the first deployed Antarctic telescope of China, CSTAR has successfully fulfilled its preset missions. Its results have proved that the astronomical observing conditions of Dome A are excellent for time-domain observations, and the scientific significance of Dome A. With more Chinese astronomical programs (e.g. AST3: Cui et al. 2008) put to work, astronomy on Dome A is about to bloom.

\acknowledgments
We would like to thank the anonymous referee for his/her very constructive and insightful comments on our original paper.
We thank Zeyang Meng and Ming Yang for their valuable advice on eliminating the false positives.
We would also like to thank Jifeng Liu and Xin Cheng for their insightful opinions on stellar flares.
We thank Xiaojia Zhang for her kind help on improving the visual effect of Figure \ref{fig5}.
This research is supported by
the National Basic Research Program of China (Nos. 2013CB834900, 2013CB834902, 2014CB845704)
the National Natural Science Foundation of China under grant Nos. 10925313, 11333002, 11433005 and 11503009;
the Strategic Priority Research Program-The Emergence of Cosmological Structures of the Chinese Academy of Sciences (Grant No. XDB09000000).

\clearpage
\begin{turnpage}
\setlength{\tabcolsep}{2.0pt}
\begin{deluxetable}{cccccccccccccccccc}
\tablecaption{Confirmed Optical Flares in CSTAR Field}
\tablewidth{0pt}
\tabletypesize{\footnotesize}
\tablehead{
\colhead{2MASS ID}     &  \colhead{$i$}      & \colhead{Duration}  &  \colhead{Amplitude}  &\colhead{Skewness}  &\colhead{$T_{\rm 0}$}      &  \colhead{$J$}        &   \colhead{$H$}        &  \colhead{$K$}     & \colhead{$B$}&\colhead{$V$}   &  \colhead{$\mu_{\rm \alpha}$}       &  \colhead{$\mu_{\rm \delta}$}      & \colhead{$T_{\rm eff}$}    & \colhead{Sp.Type}   & \colhead{Lum.Class} & \colhead{$Rosat$ ID} & \colhead{Var}  \\
\colhead{2MASS+J}         &  \colhead{(mag)}  & \colhead{(minutes)}  &  \colhead{}         & \colhead{} &\colhead{(2454500.0+)}  &\colhead{(mag)}  &  \colhead{(mag)}  & \colhead{(mag)}  &  \colhead{(mag)}&  \colhead{(mag)} & \colhead{(${\rm mas\,yr^{-1}}$ )}      & \colhead{(${\rm mas\,yr^{-1}}$ )}     & \colhead{(K)}        & \colhead{}     & \colhead{}          & \colhead{1RXS+J}    &   \\
}
\startdata
024523.37-872730.50 &  9.930 & 11.69 & 3.6$\%$ & 2.66 & 105.87 & 8.335 & 7.564 & 7.411 & 12.72 & 10.94 & 17.7 & -11.8 & 3754$^{*}$ &K5$^{*}$& III$^{*}$ &...& ...\\    %43.234
030115.76-875910.12 &  10.901 & 24.87 & 2.4$\%$ & 3.98 & 120.08 & 9.370 & 8.703 & 8.559 & 12.68 & 11.82 & 11.0 & 0.4 & 4557$^{*}$ & K3$^{*}$ & III$^{*}$ &...&...\\    %36.567
031821.47-881506.29 &  11.899 & 25.06 & 7.7$\%$ & 3.93 & 99.16 & 11.032 & 10.778 & 10.678 & 12.720 & 12.111 & 10.8 & 13.1 & 5523$^{*}$ & G5$^{*}$ &V$^{*}$ &... &...\\     %29.217
061434.95-880143.38 &  11.984 & 24.87 & 6.7$\%$ & 4.07 & 131.96 & 11.169 & 10.810 & 10.659 & 13.60 & 12.51 & 50.6 & -27.2 & 4536 &K4& Dwarf&... &...\\     %8.166
070257.77-880817.21 &  12.487 & 19.97 & 19.7$\%$ & 3.06 & 89.20 & 11.205 & 10.658 & 10.515 & 14.357 & 13.261 & -6.0 & 16.2 & 4523 &K4&Dwarf&... &...\\     %2.990
100549.80-890513.36 &  12.619 & 25.02 & 19.2$\%$ & 4.18 & 87.10 & 11.449 & 10.934 & 10.845 & 13.62 & 13.22 & 7.1 & 11.6 & 6880 &F4&Dwarf&... &...\\        %5.691
124144.03-875828.48 &  11.394 & 260.08 & 10.5$\%$ & 5.50 & 131.52 & 10.090 & 9.599 & 9.444 & 12.81 & 11.95 & -15.3 & 1.4 & 4729$^{*}$ & K0$^{*}$ &III$^{*}$ &... & ELL$\rightarrow$UV Ceti \\   %19.215
124144.03-875828.48 &  11.394 & 24.17 & 6.4$\%$ & 3.00 & 131.86 & 10.090 & 9.599 & 9.444 & 12.81 & 11.95 & -15.3 & 1.4 & 4729$^{*}$ & K0$^{*}$ & III$^{*}$&... & ELL$\rightarrow$UV Ceti \\   %19.215
142234.94-871930.54 & 10.952 & 12.09 & 6.6$\%$ & 3.00 & 98.11 & 9.450 & 8.804 & 8.608 & 13.181 & 11.889 & 3.9 & 1.2 & 4135 &K6&Giant&...&...\\    %29.831
174728.65-884609.43 &  11.848 & 29.98 & 26.5$\%$ & 4.88 & 74.27 & 9.992 & 9.386 & 9.072 & 13.740 & 12.450 & -4.3 & -26.5 & 4139 & M3.5$^{*}$ &Dwarf & 174721.0-884615 & BY$\rightarrow$UV Ceti \\    %4.213
174728.65-884609.43 &  11.848 & 15.11 & 15.0$\%$ & 1.96 & 97.24 & 9.992 & 9.386 & 9.072 & 13.740 & 12.450 & -4.3 & -26.5 & 4139 & M3.5$^{*}$ &Dwarf & 174721.0-884615 & BY$\rightarrow$UV Ceti \\    %4.213
194153.46-873535.30 & 11.539 & 18.39 & 6.5$\%$ & 4.52 & 120.55 & 9.697 & 8.866 & 8.659 & 13.50 & 12.89 & 4.3 & -17.4 & 5928 &G0& Giant&...&...\\      %10.409
195407.67-874359.52 &  10.110 & 20.08 & 4.1$\%$ & 2.96 & 115.59 & 9.300 & 9.051 & 9.025 & 10.739 & 10.339 & 26.7 & -46.7 & 6150$^{*}$ & F8$^{*}$ &V$^{*}$ &...&...\\    %11.302
202819.01-872820.05 &  7.798 & 15.08 & 1.0$\%$ & 3.99 & 80.18 & 6.863 & 6.573 & 6.475 & 8.79 & 8.08 & 169.9 & -57.8 & 5707$^{*}$ & G2$^{*}$ &V$^{*}$ &...&...\\    %18.101
211308.99-875632.28 &  9.949 & 15.09 & 2.1$\%$ & 3.99 & 104.30 & 8.338 & 7.640 & 7.452 & 12.34 & 11.03 & -6.0 & -1.1 & 4362$^{*}$ & K4$^{*}$ &V$^{*}$ &...&...\\     %14.974
\enddata

\label{tab2}
\tablenotetext{}{Flare events are arranged with increasing right ascension coordinates.
       The column descriptions are as follows: 
       Col. (1): 2MASS identifier of variability. 
       Col. (2): Median $i$ apparent magnitude.
       Col. (3): Duration of flare event.
       Col. (4): Amplitude of flare event.
       Col. (5): Skewness of flare event.
       Col. (6): Flare event starting date.
       Col. (7-11): $JHKBV$  magnitudes of flaring sources from the 2MASS catalog.
       Col. (12,13): Proper motion of flaring sources in right ascension and declination from the PPMXL catalog.
       Col. (14): Effective temperature from VizieR database (if available) and our calculation. 
       Col. (15): Spectral type of flaring source from VizieR database (if available) and our calculation.
       Col. (16): Luminosity Class of flaring source from VizieR database (if available) and our calculation.
       Col. (17): $Rosat$ identifier of flaring source, if available. 
       Col. (18): Flaring variable reclassification: 
                       BY, BY Draconis-type variables;
                       ELL, Rotating ellipsoidal variables.
Stellar properties taken from the VizieR database are marked with asterisks.}
\end{deluxetable}
\end{turnpage}
\clearpage

\global\pdfpageattr\expandafter{\the\pdfpageattr/Rotate 90}

\end{document}